# A study on rare-earth Laves phases for magnetocaloric liquefaction of hydrogen

Wei Liu[1, *], Eduard Bykov[2, 3], Sergey Taskaev[4, 6], Mikhail Bogush[4, 6], Vladimir Khovaylo[5], Nuno Fortunato[1], Alex Aubert[1], Hongbin Zhang[1], Tino Gottschall[2, *], Jochen Wosnitza[2,3], Franziska Scheibel[1], Konstantin Skokov[1], Oliver Gutfleisch[1]

[1]Institute of Materials Science, Technical University of Darmstadt, 64287 Darmstadt, Germany
[2]Dresden High Magnetic Field Laboratory (HLD-EMFL) and Würzburg-Dresden Cluster of Excellence ct.qmat, Helmholtz-Zentrum Dresden-Rossendorf, 01328 Dresden, Germany
[3]Institute of Solid State and Materials Physics, Technische Universität Dresden, 01062 Dresden, Germany
[4]Faculty of Physics, Chelyabinsk State University, 454001 Chelyabinsk, Russia
[5]Department of Functional Nanosystems and High-Temperature Materials, National University of Science and Technology "MISiS", 119991 Moscow, Russia
[6]Functional Materials Laboratory, NRU South Ural State University, 454080 Chelyabinsk, Russia

*Authors to whom correspondence should be addressed:
wei.liu@tu-darmstadt.de
Alarich-Weiss-str. 16, 64287 Darmstadt
t.gottschall@hzdr.de
Bautzner Landstr. 400, 01328 Dresden

We are witnessing a great transition towards a society powered by renewable energies to meet the ever-stringent climate target. Hydrogen, as an energy carrier, will play a key role in building a climate-neutral society. Although liquid hydrogen is essential for hydrogen storage and transportation, liquefying hydrogen is costly with the conventional methods based on Joule-Thomas effect. As an emerging technology which is potentially more efficient, magnetocaloric hydrogen liquefaction can be a "game-changer". In this work, we have investigated the rare-earth-based Laves phases $R$Al$_2$ and $R$Ni$_2$ for magnetocaloric hydrogen liquefaction. We have noticed an unaddressed feature that the magnetocaloric effect of second-order magnetocaloric materials can become "giant" near the hydrogen boiling point. This feature indicates strong correlations, down to the boiling point of hydrogen, among the three important quantities of the magnetocaloric effect: the maximum magnetic entropy change $\Delta S_m^{max}$, the maximum adiabatic temperature change $\Delta T_{ad}^{max}$, and the Curie temperature $T_C$. Via a comprehensive literature review, we interpret the correlations for a rare-earth intermetallic series as two trends: (1) $\Delta S_m^{max}$ increases with decreasing $T_C$; (2) $\Delta T_{ad}^{max}$ decreases near room temperature with decreasing $T_C$ but increases at cryogenic temperatures. Moreover, we have developed a mean-field approach to describe these two trends theoretically. The dependence of $\Delta S_m^{max}$ and $\Delta T_{ad}^{max}$ on $T_C$ revealed in this work helps researchers quickly anticipate the magnetocaloric performance of rare-earth-based compounds, guiding material design and accelerating the discoveries of magnetocaloric materials for hydrogen liquefaction.

## 1. Introduction

The European Union has set a target of being climate-neutral by 2050 – an economy with net-zero greenhouse-gas emissions [1,2]. In this paper we focus on the potential of a carbon-neutral "hydrogen society" [3–5], where hydrogen plays a key role by acting as the vector between energy generation and energy utilization. Produced from renewables, green hydrogen is one of the cleanest fuels, ideally releasing no carbon and no pollutants into the atmosphere [4,6,7]. **Figure 1** is a schematic picturing how the carbon-neutral "hydrogen society" works, providing that the hydrogen fuel can be made to operate reliably and at a price that society is willing to pay.

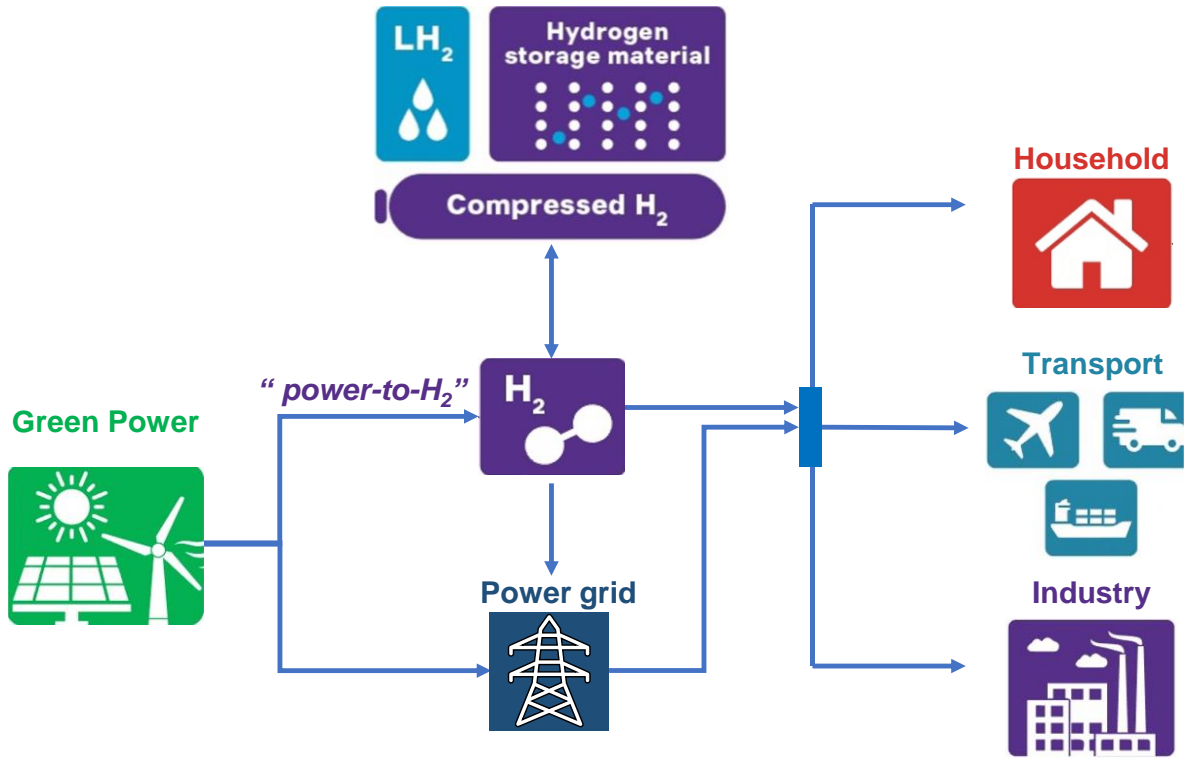

**Figure 1.** A carbon-neutral hydrogen society: the hydrogen is produced by renewables such as solar and wind power; it can be stored and transported as compressed, liquefied, or interstially maintained hydrogen. The hydrogen can later be released as a fuel for households, transport, and industry. In addition, the hydrogen energy can be converted back into electricity and integrated into the power grid.

Hydrogen can be stored and transported efficiently in its liquid state [4,8–11]. However, conventional liquefaction based on the Joule-Thomas effect is expensive [12,13], and becomes increasingly inefficient as we approach cryogenic temperatures [14,15]. The magnetocaloric liquefaction technology, with the use of superconducting coils [16–20] or permanent magnets [21,22] at cryogenic temperatures, is ideally more efficient [22–31]. This leads to a recent growing interest in liquifying hydrogen and other industrial gases with magnetic cooling [32–38]. It is worth mentioning that a practical active magnetic regenerator for hydrogen liquefaction using superconducting magnets that generate fields up to 5 T has been built by Kamiya *et al.* recently [20]; and that Feng *et al.* demonstrated the feasibility of using 1 T field generated by permanent magnets by modeling [22].In terms of the cost, permanent magnets have an advantage in low fields [22]. Although more expensive, the superconducting magnets have the advantage of generating high fields. Which type of magnets to choose for magnetocaloric hydrogen liquefaction on industrial scale needs further studying, and certainly the maximum magnetocaloric effect of the materials has an influence on the decision.

Discovered by Weiss and Picard [39], the magnetocaloric effect is the warming or cooling of a magnetic material when exposed to a magnetic field. This effect is best characterized by two physical quantities, i.e., the magnetic entropy change $\Delta S_m$ and the adiabatic temperature change $\Delta T_{ad}$ that indicate the potential cooling power of a magnetocaloric material [40]. Since $\Delta S_m$ and $\Delta T_{ad}$ usually peak at Curie temperature $T_C$ [41] for second-order magnetocaloric materials, we assume that $\Delta S_m(T_C)$ and $\Delta T_{ad}(T_C)$ are the maximum magnetic and adiabatic temperature changes ($\Delta S_m^{max}$ and $\Delta T_{ad}^{max}$). In other words, $T_C$ is the best-suited working temperature of a magnetocaloric material.

In the case of magnetocaloric liquefaction of hydrogen, the gaseous hydrogen needs to be cooled down from around room temperature to its boiling point of about 20 K [42]. This is a large temperature range. If the gaseous hydrogen is pre-cooled by liquid nitrogen, the working temperature range of magnetocaloric hydrogen liquefaction is reduced to 77 K~ 20 K [43]. This is still a large temperature range. The large working temperature ranges required by magnetocaloric hydrogen liquefaction really put a challenge on the material design because it is usually difficult for a single magnetocaloric material to cover such large temperature ranges. A combination of different magnetocaloric materials with Curie temperatures distributed appropriately within the working temperature range is necessary [17]. In this context, rare-earth-based alloys are promising candidates for magnetocaloric hydrogen liquefaction, not only due to the large $\Delta S_m^{max}$ and $\Delta T_{ad}^{max}$ enabled by the huge total magnetic moments [44–46], but also due to the tunable $T_C$ that covers the cryogenic temperatures [47–49]. Indeed, the $T_C$ of the rare-earth alloys can easily be tuned by substitution of different rare-earth elements to approach the hydrogen boiling point of 20 K [50–52]. Among the rare-earth-based magnetocaloric materials, some materials such as $Eu_4PdMg$ [53], $ErZn_2$/ErZn composite [54], and amorphous $Er_{0.2}Gd_{0.2}Ho_{0.2}Co_{0.2}Cu_{0.2}$ [55] showing a "table-like" maximum magnetocaloric effect over a large temperature span. Considering that the maximum magnetocaloric effects need to be considered for designing, for example, a composite with a "table-like" magnetocaloric effect [56], studying the relationships among $\Delta S_m^{max}$, $\Delta T_{ad}^{max}$, and the ordering temperature (in this work we focus on $T_C$) is meaningful.

From our study on rare-earth Laves phases $R$Al$_2$ and $R$Ni$_2$ ($R$: Gd, Tb, Dy, Ho, Er), we discovered a noticeable, but unaddressed feature for the rare-earth magnetocaloric materials: maximum magnetocaloric effect is much stronger near the boiling point of hydrogen. In this work, we achieved a fully reversible value of $\Delta T_{ad}^{max}$ directly measured under pulsed fields in HoNi$_2$: 5.7 K at 15 K in fields of 2 T. In even lower temperature, even larger values of $\Delta T_{ad}^{max}$ indirectly measured were reported: 6.2 K at around 4.4 K in fields of 1 T for TmCoSi [57] and 8.4 K at around 13 K in fields of 2 T for ErCr$_2$Si$_2$ [58]. Moreover, many second-order magnetocaloric materials also have "giant" values of $\Delta S_m^{max}$ near 20 K: ErAl$_2$ [59], HoB$_2$ [60], and TmGa [35] which have a magnetic entropy change being larger than or close to the first-order giant magnetocaloric material ErCo$_2$ [61], whereas it is not the case near room temperature where second-order magnetocaloric materials can barely reach the value of ErCo$_2$. In even lower temperature than 20 K, even larger values of $\Delta S_m^{max}$ were reported for the second-order magnetocaloric material GdF$_3$: 45.5 J kg$^{-1}$ K$^{-1}$ in fields of 2 T and 67.1 J kg$^{-1}$ K$^{-1}$ in fields of 5 T at around 2 K [62]. The "giant" values of $\Delta S_m^{max}$ and $\Delta T_{ad}^{max}$, which are usually attributed to the first-order magnetic phase transition, are discovered in the second-order magnetocaloric materials such as HoNi$_2$ and ErAl$_2$ with a Curie temperature close to the boiling point of hydrogen. This indicates that there is strong correlations among $\Delta S_m^{max}$, $\Delta T_{ad}^{max}$ and $T_C$.

In this article, we raise a question with a great importance for material design: how do $\Delta S_m^{max}$ and $\Delta T_{ad}^{max}$ correlate with $T_C$, particularly in the vicinity of hydrogen boiling point? In the present work, we use the heavy rare-earth -based Laves phases $R$Al$_2$ and $R$Ni$_2$ as an example to study the correlations among $\Delta S_m^{max}$, $\Delta T_{ad}^{max}$ and $T_C$ because of the chemical and physical similarities of these heavy rare-earth elements [63], and their Curie temperatures that covers a large temperature range down to the hydrogen boiling point. Although heavy rare-earth elements such as Tb and Ho are not abundant in Earth's upper continental crust [63], which is a challenge for their alloys to be used for magnetocaloric hydrogen liquefaction on an industrial scale, this way of studying the correlations among $\Delta S_m^{max}$, $\Delta T_{ad}^{max}$ and $T_C$ will benefit the research on the magnetocaloric alloys of their light rare-earth counterparts such as Nd and Pr, which are more abundant than the heavy rare-earth elements [63].

To answer the above proposed question, we divide our study into three. First, we look at the rare-earth-based Laves phases $R$Al$_2$ and $R$Ni$_2$, where we observe a noticeable, but unaddressed feature that maximum magnetocaloric effect is much stronger near the boiling point of hydrogen. Next, we summarize two trends for a rare-earth-based intermetallic series via the experimental work on the rare-earth-based Laves phases and a comprehensive literature review: (1) $\Delta S_m^{max}$ increases with decreasing $T_C$; (2) $\Delta T_{ad}^{max}$ decreases near room temperature with decreasing $T_C$ but increases at cryogenic temperatures. In the third part we develop a mean-field approach, with $R$Ni$_2$ and $R$Al$_2$ as examples, that provides a theoretical explanation for these two trends. The dependence of $\Delta S_m^{max}$ and $\Delta T_{ad}^{max}$ on $T_C$ described in this work not only explains why a second-order magnetocaloric material can have a "giant" magnetocaloric effect near hydrogen boiling point, but also helps us quickly anticipate the magnetocaloric performance of rare-earth-based compounds, guiding material design and accelerating the discoveries of magnetocaloric materials for hydrogen liquefaction.

## 2. Experimental case study on $R$Al$_2$ and $R$Ni$_2$

We arc melted the HoAl$_2$ and GdNi$_2$ samples as well as the other $R$Ni$_2$ and $R$Al$_2$ samples that had 3% excess of rare-earth elements. X-ray powder diffraction data were collected on a powder diffractometer (Stadi P, Stoe&Cie GmbH) equipped with a Ge[111]-monochromator using Mo-K$_\alpha$ radiation (λ = 0.70930 Å) in the Debye-Scherrer geometry. The data were evaluated with Rietveld refinements using the FullProf package [64]. The analyses revealed that the phase fractions of HoAl$_2$ and GdNi$_2$ were close to 90%, while those of the 3%-excess samples $R$Al$_2$ and $R$Ni$_2$ were close to 100% and 95% respectively. We further confirmed the sufficient quality of our samples by backscatter electron (BSE) imaging and energy-dispersive X-ray spectroscopy (EDS) using a Tescan Vega 3 scanning electron microscope (SEM) equipped with an EDAX Octane Plus detector. Magnetizations versus applied field as a function of temperature were measured with a Physical Properties Measurement System (PPMS) from Quantum Design in fields up to 5 T. Direct measurements of the adiabatic temperature change in pulsed fields up to 20 T were made at the Dresden High Magnetic Field Laboratory. Each sample was cut into two flat pieces with a dimension of around 3 mm x 3 mm x 2 mm. A thin type-E thermocouple with a wire thickness of 25 $\mu m$ was fixed between these pieces by a little amount of silver epoxy. After the samples were fixed on the sample holder, the insert was evacuated to a high vacuum of $10^{-5}\ mbar$. The time to reach the maximum field was always 13 ms. More detailed descriptions of the measurements in pulsed fields can be found in references [45,46,65,66].

**Figure 2** shows the magnetic moment as a function of temperature for the Laves phases materials $R$Ni$_2$ and $R$Al$_2$ in a magnetic field of 5 T. This is a field which can be easily generated by superconducting magnets, and in which many magnetocaloric materials are studied. The heavy rare-earth elements from Gd to Er possess a large magnetic moment due to the partially filled 4$f$ shells, resulting in a high saturation magnetization in $R$Ni$_2$ and $R$Al$_2$. For all the investigated materials, the magnetization around 10 K is always above 6 $\mu_B$ per rare-earth atom, higher than the 5.9 $\mu_B$ of Mn$^{2+}$ and Fe$^{3+}$.

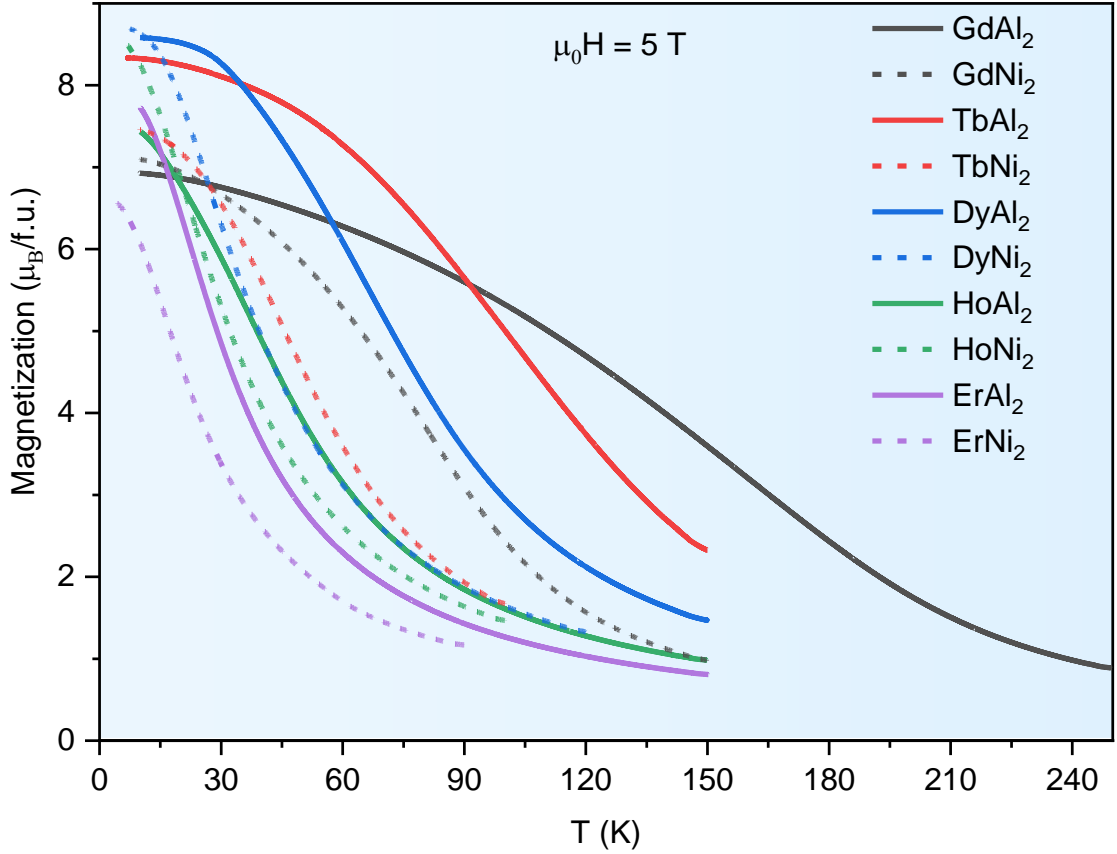

**Figure 2.** Magnetic moment of the $R\text{Al}_2$ and $R\text{Ni}_2$ Laves phases as a function of temperature in fields of 5 T.

By applying the Curie-Weiss law we can determine the effective magnetic moment as:

$$\mu_{eff} = \frac{1}{\mu_B}\sqrt{\frac{3k_B M_R}{N_A \alpha}}, \qquad (1)$$

where $\mu_B$ is the Bohr magneton, $k_B$ is the Boltzmann constant, $M_R$ is the molecular mass, $N_A$ is the Avogadro constant, and $\alpha$ is the slope of $\mu_0\chi^{-1}$ vs. $T$, where $\mu_0$ is the vacuum permeability and $\chi$ is the magnetic susceptibility. In **Figure 3** we plotted $\mu_0\chi^{-1}$ vs. $T$ in fields of 1 T, with dashed lines denoting the Curie-Weiss fits. The intercepts with the *x*-axis represent the paramagnetic Curie temperatures. The effective magnetic moments $\mu_{eff}$ of the compounds as well as $\mu_{eff}$ of the corresponding free rare-earth ions are summarized together with $T_C$ in **Table 1**. The effective magnetic moments of $R\text{Ni}_2$ and $R\text{Al}_2$ range from about 8 to 11 $\mu_B$/f. u. with an average of 9.85 $\mu_B$/f. u., putting them close to $\mu_{eff}$ of the corresponding free rare-earth ions. This confirms the large and similar magnetic moments of these heavy rare-earth compounds.

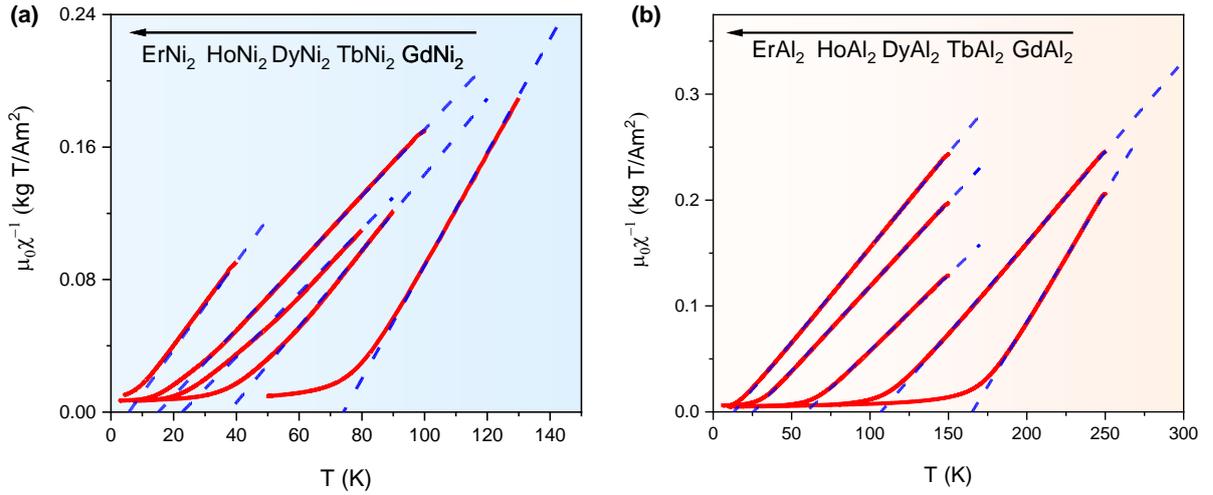

**Figure 3.** $\mu_0\chi^{-1}$ as a function of temperature in fields of 1 T with Curie-Weiss fits shown as red dashed lines for (a) $R\text{Ni}_2$ and (b) $R\text{Al}_2$.

**Table 1.** Effective magnetic moments of the $R\text{Al}_2$ and $R\text{Ni}_2$ Laves phases calculated using the Curie-Weiss law and the effective magnetic moments of rare-earth ions taken from reference [52].

| Compounds | $\mu_{eff}$ in $R\text{Al}_2$ and $R\text{Ni}_2$ ($\mu_B$/f. u.) | $\mu_{eff}$ of free ions ($\mu_B$) | $T_C$ (K) |
|---|---|---|---|
| GdAl$_2$ | 8.32 | 8.9 | 165 |
| GdNi$_2$ | 8.01 | | 74 |
| TbAl$_2$ | 9.98 | 9.8 | 107 |
| TbNi$_2$ | 9.82 | | 38 |
| DyAl$_2$ | 11.12 | 10.6 | 60 |
| DyNi$_2$ | 10.74 | | 23 |
| HoAl$_2$ | 10.46 | 10.4 | 26 |
| HoNi$_2$ | 10.60 | | 15 |
| ErAl$_2$ | 9.97 | 9.5 | 13 |
| ErNi$_2$ | 9.47 | | 6 |

**Figure 4** shows the magnetic entropy change in fields of 5 T and 2 T, and the adiabatic temperature change measured directly in pulsed fields up to 20 T for $R\text{Ni}_2$ and $R\text{Al}_2$ (except ErNi$_2$ due to the low sensitivity of the thermal couple below 10 K). The magnetic entropy change was calculated using the Maxwell relation and a numerical integration of $\Delta S_m = \int_0^H \left(\frac{\partial M}{\partial T}\right)_H dH$ ($H$ is the magnetic field and $M$ is the magnetization) with a varying magnetic field up to 5 T at a given temperature, as proposed in reference [67]. For both Laves phase series we see a higher peak value at a lower $T_C$ (except for ErNi$_2$), as indicated by the gray arrows. The $\Delta S_m^{max}$ of ErAl$_2$ with a $T_C$ of 13 K reaches a value close to 40 J kg$^{-1}$ K$^{-1}$ in fields of 5 T. This value is five times larger than that of GdAl$_2$, with a $T_C$ of 165 K. The $\Delta T_{ad}^{max}$ of HoNi$_2$ with a $T_C$ of 15 K has a fully reversible value of 10.6 K in fields of 5 T. This value is more than twice larger than that of GdNi$_2$, with a $T_C$ of 74 K.

These observations point to a close correlation between the maximum magnetocaloric effect and the $T_C$, while, at least for $R\text{Ni}_2$ (excluding ErNi$_2$) and $R\text{Al}_2$ series, a larger magnetocaloric effect is always observed at lower temperatures. But is this just coincidence, or a genuine feature of rare-earth-based magnetocaloric materials? In the search for answer, in the next section we review the literature on rare-earth-based magnetocaloric materials from near-room temperature down to 20 K.

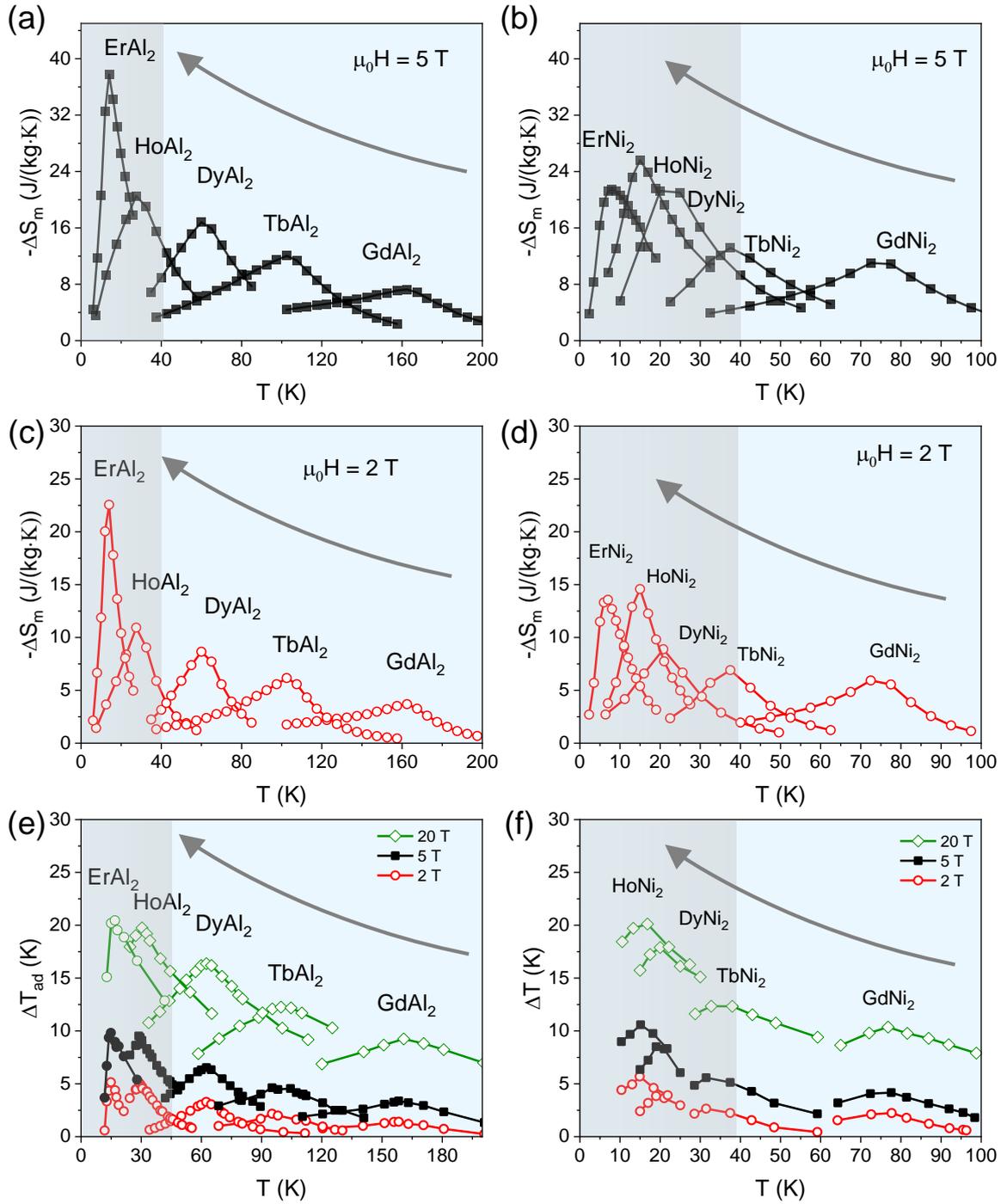

**Figure 4.** (a) (b) (c) (d) Magnetic entropy changes in fields of 5 and 2 T for the $R$Al$_2$ and $R$Ni$_2$ Laves phases, and (e) (f) direct measurements of the adiabatic temperature changes in pulsed fields of 2, 5, and 20 T, excluding ErNi$_2$.

## 3. Summarizing two trends from literature review

**Figure 5** shows reported values of $\Delta S_m(T_C)$ for rare-earth-based materials [60,61,68–111] together with our results in fields of 5 T. There is an additional table in the Supplementary that contains the values for this graph and the corresponding references. The distribution of Curie temperatures for a single rare-earth-intermetallic series is similar to $R$Al$_2$ and $R$Ni$_2$, i.e., $T_C$ decreases with increasing atomic number from Gd to Tm. This is due to the reducing de Gennes factors from Gd to Tm [50–52]. It should be emphasized that upon evaluating the magnetocaloric effect, the reversibility caused by thermal hysteresis for the first-order magnetocaloric materials, namely ErCo$_2$, HoCo$_2$, DyCo$_2$, Dy$_5$Si$_4$, Ho$_5$Si$_4$, Er$_5$Si$_4$ in the reviewed data, should be taken into consideration. The reversibility of these compounds need to be studied carefully since there is a lack of sufficient articles addressing this point for these compounds.

Large magnetic entropy changes occur below 140 K, exceeding that of the benchmark polycrystalline Gd with a value of about 10 J kg$^{-1}$ K$^{-1}$ [98]. "Giant" values are found near 20 K (indicated by the gray shadow), where the magnetic entropy change of HoB$_2$ reaches 40 J kg$^{-1}$ K$^{-1}$ in fields of 5 T [60], four times larger than that of Gd [98], and better than that of ErCo$_2$ [61], which is a first-order giant magnetocaloric material. Like with the $R$Al$_2$ and $R$Ni$_2$ series, an increasing $\Delta S_m^{max}$ is measured as the $T_C$ decreases for other rare-earth-based intermetallic series, although there are exceptions, such as TbFeSi, TbNi, and Ho$_3$CoNi. Based on these empirical findings in the literature we propose the first trend: the maximum magnetic entropy change of an ideal rare-earth-based intermetallic series increases as the Curie temperature decreases.

In the inset in **Figure 5**, we also plot the maximum magnetic entropy changes of some of the light rare-earth-based magnetocaloric materials collected from literature [112–117], where an increasing trend with decreasing Curie temperature can be also observed. Things are more complicated for the light rare-earth-based compounds because it is difficult to build a material family like their heavy rare-earth counterparts where the distribution of the Curie temperatures can cover a large temperature range. Generally, magnetocaloric effect in the light rare-earth-based alloys for hydrogen liquefaction are less studied than that of their heavy rare-earth counterparts. More studies on the light rare-earth-based compounds are needed considering that the light rare-earth elements such as Ce, Nd, and Pr are much more abundant than heavy rare-earth elements such as Tb and Ho [63], although their alloys generally show a weaker magnetocaloric effect than those of their heavy rare-earth counterparts due to their smaller magnetic moments. Balancing the performance and the costs is still an open question which needs to be investigated. Studying the correlations among $\Delta S_m^{max}$, $\Delta T_{ad}^{max}$, and $T_C$ will contribute to this question.

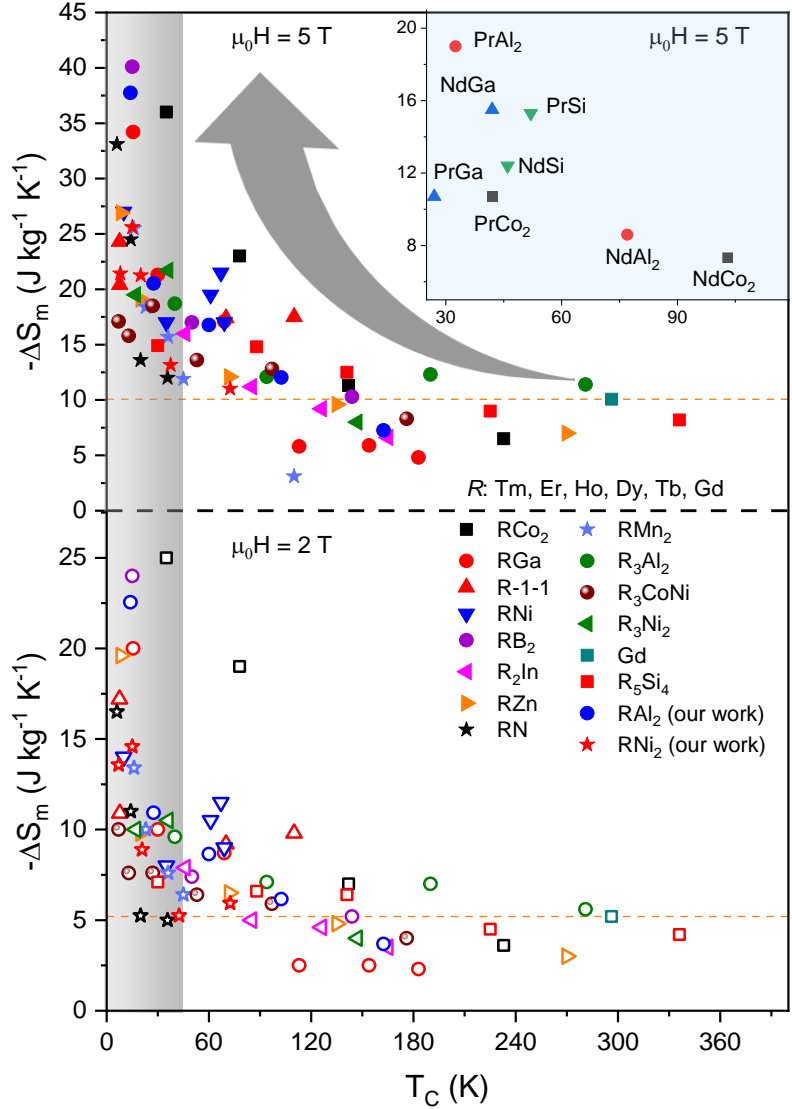

**Figure 5.** Magnetic entropy changes in fields of 5 and 2 T as a function of Curie temperature for heavy rare-earth-based magnetocaloric matelals, the inset shows the mangetic entropy changes in 5 T for the light rare-earth-based magnetocaloric materials. Data for $R$Al$_2$ and $R$Ni$_2$ are from our work, the other data are collected from literature [60,61,68–117]. The horizontal orange dashed lines mark the magnetic entropy change of Gd, the gray shadow indicates the region in the vicinity of hydrogen boiling point.

In contrast to the numerous data on the magnetic entropy change available in the literature, there are only limited data on the adiabatic temperature change in the range below 100 K, measured either indirectly or directly. In this study we measured the $\Delta T_{ad}$ of $R$Al$_2$ and $R$Ni$_2$ directly in pulsed magnetic fields. The results are plotted in **Figure 6** together with other published data: pure rare-earth metals [65,118], the 5:4 phases [108], the $R$Co$_2$ Laves phases [61,83,84,97,119], the 1:1:1 phases [74–76,120], the HoB$_2$ [60], the TmCoSi [57], and the ErCr$_2$Si$_2$ [58], with respect to their $T_C$ in magnetic fields of 5 and 2 T. It should be addressed that Tb and Dy are antiferromagnetic below 1 T and that TmCoSi is weakly antiferromagnetic below 0.09 T, at the temperature where they show the maximum magnetocaloric effect.

Like in **Figure 5** for $\Delta S_m^{max}$, $\Delta T_{ad}^{max}$ becomes large near hydrogen's boiling point of 20 K. For polycrystalline HoNi$_2$ measured at 15 K in a pulsed field of 2 T, $\Delta T_{ad}^{max}$ has a fully reversible value of 5.7 K, surpassing that of a Gd single crystal, about 5 K [40,65]. Towards lower temperatures, TmCoSi and ErCr$_2$Si$_2$ show a $\Delta T_{ad}^{max}$ over 8 K in fields of 2 T [57,58], surpassing that of the giant near-room-temperature first-order magnetocaloric materials $La(Fe,Si)_{13}$ and $FeRh$ with a $\Delta T_{ad}^{max}$ below 7 K [40]. However, from a $T_C$ near room temperature to a $T_C$ of about 160 K, we see that $\Delta T_{ad}^{max}$ decreases with decreasing $T_C$ for the pure rare-earth metals and the 5:4 phases. For the other series with a lower $T_C$, $\Delta T_{ad}^{max}$ increases with decreasing $T_C$ (except for TbFeSi). Based on these empirical findings we propose the second trend: the maximum adiabatic temperature change of a rare-earth-based intermetallic series decreases close to room temperature as Curie temperature decreases but increases at cryogenic temperatures.

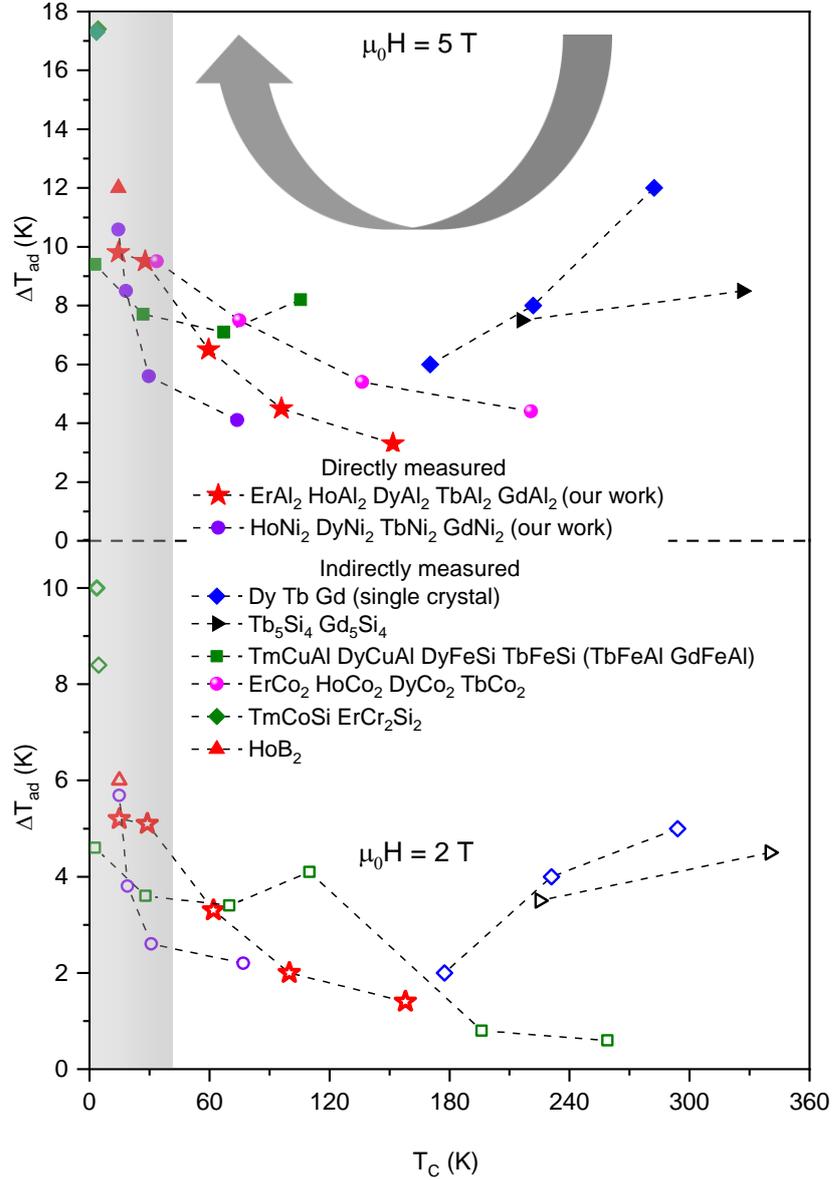

**Figure 6.** Adiabatic temperature changes of the $R$Al$_2$ and $R$Ni$_2$ Laves phases, the pure rare-earth metals [65,118], the 5:4 phase [65,108], the RCo$_2$ Laves phases [61,83,84,97,119], the 1:1:1 phase [74–76,120], the HoB$_2$ [60], the TmCoSi [57], and the ErCr$_2$Si$_2$ [58], in fields of 5 T (top) and 2 T (bottom). Note that there are no data for TbFeAl and GdFeAl in 5 T. Each sequence in the legend is listed with increasing Curie temperature from left to right. The gray shadow indicates the region in the vicinity of hydrogen boiling point.

The two trends can now help us to make a quick assessment of the magnetocaloric effect in rare-earth-based alloys. For example, HoB$_2$ which has the largest magnetic entropy change in **Figure 5**, was recently proposed by Castro *et al.*, who used machine-learning algorithms [60].

The first trend could also have been applied to anticipate the performance of HoB$_2$. Knowing that HoB$_2$ has a Curie temperature of 15 K and that TbB$_2$ [95] and DyB$_2$ [94] have large magnetic entropy changes, the first trend would suggest that HoB$_2$ should have an even larger magnetic entropy change. The two trends could also help with alloy design when we need to tune $T_C$, because they state that $\Delta S_m^{max}$ and $\Delta T_{ad}^{max}$ are both dependent on $T_C$. Moreover, these two trends suggest that there is a physical explanation as to why $\Delta S_m^{max}$ and $\Delta T_{ad}^{max}$ correlate with $T_C$.

## 4. Theoretical description via mean-field approach

The magnetocaloric effect is influenced by many factors, such as microstructure, stoichiometry, order of the phase transition, magnetic moment, and the sweep rate of the field [24,121–125]. It was shown by Gottschall *et al.* [65], Romero-Muñiz *et al.* [126], Belo *et al.* [127], and Oesterreicher *et al.* [128] that a mean-field approach can explain the magnetocaloric properties of rare-earth elements and their compounds. We have used this approach to understand how $\Delta S_m^{max}$ and $\Delta T_{ad}^{max}$ vary with $T_C$.

The magnetic entropy derived from a mean-field model is given by reference [129] as:

$$S_m = N_M k_B \left[ \ln\left(\frac{\sinh\frac{2J+1}{2J}x}{\sinh\frac{1}{2J}x}\right) - x B_J(x) \right], \quad (2)$$

where $x = \frac{\mu\mu_0 H + \frac{3J}{J+1} k_B T_C B_J(x)}{k_B T}$. $N_M$ is the number of "magnetic atoms", $B_J(x)$ is the Brillouin function, $J$ is the total angular momentum, $\mu$ is the atomic magnetic moment, $T_C$ is the Curie temperature, $\mu_0$ is the vacuum permeability, and $k_B$ is the Boltzmann constant. Gottschall *et al.* already demonstrated that this equation is applicable to a single crystal of Gd in pulsed fields up to 62 T [65]. However, it is important to note that this approach only provides good results in the paramagnetic state; it is not accurate enough to describe the ferromagnetic state with its spontaneous magnetization [65]. Nevertheless, since $\Delta S_m^{max}$ and $\Delta T_{ad}^{max}$ are located near $T_C$, where the material is only weakly ferromagnetic, this equation is useful for describing these two quantities [65].

The adiabatic temperature change $\Delta T_{ad}$ is obtained by constructing the total entropy $S(H,T)$ curves as a function of $H$ and $T$ [129]. In addition to the magnetic entropy $S_m$, we have the lattice entropy $S_l$, and the electronic entropy $S_e$ [124]. Usually, $S_e$ is small compared to $S_l$ above 10 K and can be neglected [124]. The volumetric lattice heat capacity is approximated by the Debye model as:

$$C_V = 9Nk_B \left(\frac{T}{T_D}\right)^3 \int_0^{\frac{T_D}{T}} \frac{x^4 e^x}{(e^x - 1)^2} dx, \quad (3)$$

where $T_D$ is the Debye temperature, $N$ is the total number of atoms, and $x = \hbar\omega/k_B T$ where $\hbar$ is the reduced Planck constant and $\omega$ is the circular frequency of the phonons. The integration of $C_V$ gives the lattice entropy

$$S_l = \int_0^T \frac{C_V}{T} dT = -3Nk_B \left[1 - \exp\left(-\frac{T_D}{T}\right)\right] + 12Nk_B \left(\frac{T}{T_D}\right)^3 \int_0^{\frac{T_D}{T}} \frac{x^3}{\exp(x) - 1} dx. \quad (4)$$

The total entropy $S(H,T)$ curves are constructed by summing $S_m$ and $S_l$. The adiabatic temperature change is then given by

$$\Delta T_{ad}(T_I, H_I \to H_F) = T_F(S_F, H_F) - T_I(S_I, H_I), \quad (5)$$

where $T_I$ and $T_F$ are the initial and final temperatures, $H_I$ and $H_F$ are the initial and final magnetic fields, and $S_I$ and $S_F$ are the initial and final total entropies with $S_I = S_F$ [129].

The results for $\Delta S_m$ calculated with equation (2) and $\Delta T_{ad}$ calculated with equation (5) in fields of 5 T are shown in **Figure 7**. We use the parameters $\mu = 10\ \mu_B$/f. u. and $J = 7.5$, which correspond to DyAl$_2$. The volumetric lattice heat capacities of $R$Al$_2$ can be reasonably assumed to lie in between that of LaAl$_2$ with a Debye temperature of 352 K, and that of LuAl$_2$ with a Debye temperature of 384 K [130–133]. **Figure 7** (b) and (c) show the calculated $\Delta T_{ad}$ with $T_D = 352$ and $384\ K$ respectively. In agreement with the two trends, $\Delta S_m^{max}$ increases with decreasing $T_C$, and $\Delta T_{ad}^{max}$ decreases first from near room temperature but then increases towards a lower $T_C$. Both quantities exhibit large values near 20 K, as highlighted in the shadows in **Figure 7**.

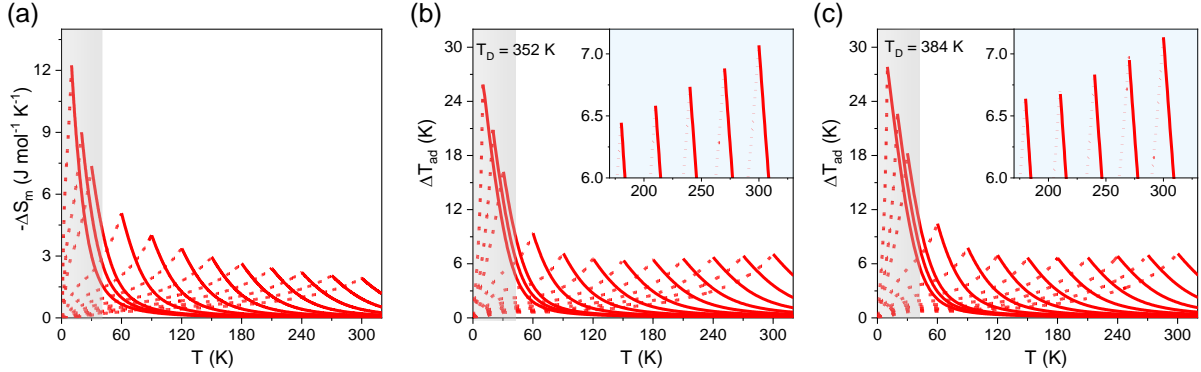

**Figure 7.** Calculations using mean-field theory of (a) the magnetic entropy change as a function of temperature in fields of 5 T for different $T_C$, (b) (c) adiabatic temperature change as a function of temperature in fields of 5 T for different $T_C$ with $T_D = 352$ and 384 K. The gray shadows indicate the regions in the vicinity of hydrogen boiling temperature and the insets in (b) and (c) shows the enlarged region in the vicinity of room temperature.

This mean-field approach can be approximated to a simple form that directly links $\Delta S_m^{max}$ and $\Delta T_{ad}^{max}$ to $T_C$. Regarding $\Delta S_m$, in 1984 Oesterreicher *et al.* found an approximate relationship between $\Delta S_m^{max}$, $T_C$ and $H$, i.e., $\Delta S_m^{max} \propto (H/T_C)^{2/3}$, by using the Taylor expansion of the Brillouin function, and they considered the cases near room temperature and above [128]. Combining Landau theory and the mean-field model, in 2012 Belo *et al.* demonstrated that many systems, especially the heavy rare-earth-based systems, fit to the relationship $\Delta S_m(T_C) \propto T_C^{-2/3}$ [127]. **Figure 8** shows the molar $\Delta S_m^{max}$ as a function of $T_C^{-2/3}$ using the calculated data from **Figure 7** (a) and the experimental data from **Figure 4**. The data from the mean-field approach demonstrate that the linear relationship between $\Delta S_m^{max}$ and $T_C^{-2/3}$ holds, at least down to 20 K in fields of 5 T, as indicated by the violet line. Although the experimental values for the polycrystalline samples lie somewhat below the theoretical values, the $T_C^{-2/3}$ relationship remains valid down to ErAl$_2$, with a Curie temperature close to 13 K. However, ErNi$_2$ with a $T_c$ of about 6 K deviates from the $T_C^{-2/3}$ line significantly. Although its effective magnetic moment is calculated to be 9.47 $\mu_B$, which is close to the average of 9.85 $\mu_B$, this deviation suggests that there are specific factors beyond this mean-field approach below 10 K. The abnormal variation trend of $\Delta S_m^{max}$ among the $R$Ni$_2$ series shown in the present work is in accordance with Ref. [132], in which the influence of the crystalline electric field on the magnetocaloric effect is taken into consideration. We suspect that the crystalline electric field has a large influence on the magnetocaloric effect of the rare-earth-based materials below 10 K. This, however, needs more study both theoretically and experimentally. To sum up our mean-field approach on $\Delta S_m^{max}$, the function of $\Delta S_m(T_C) \propto T_C^{-2/3}$ approximating equation (2) is a negative exponential function that undergoes an increasingly sharper increase towards an ever lower $T_C$. This gives us a mathematical explanation for the empirical finding that $\Delta S_m$ exhibits large values around 20 K.

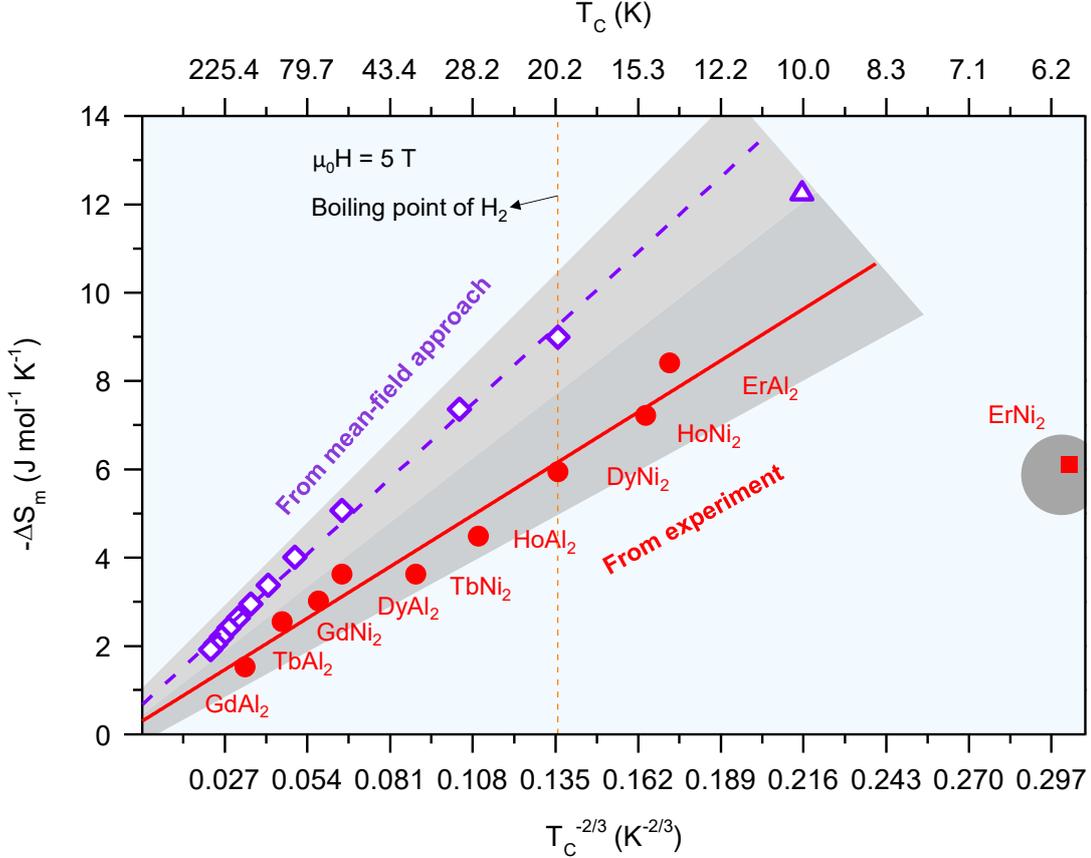

**Figure 8.** Experimental (red circles and square in the dark gray shadows) and mean-field-theory data (violet open diamonds and open triangle in the light gray shadows) of the magnetic entropy change in $R$Al$_2$ and $R$Ni$_2$ as a function of $T_C^{-2/3}$; the violet dashed line is the linear fitting of the calculated data and the red line is the linear fitting of the experimental data. The top axis shows the corresponding Curie temperature $T_C$, the vertical orange dashed line indicates the boiling point of hydrogen.

As for the adiabatic temperature change, the dependence between $\Delta T_{ad}^{max}$ and $T_C$ is not as well studied as that between $\Delta S_m^{max}$ and $T_C$. In 2011, Kuz'min *et al.* revealed that $\Delta T_{ad}^{max} \propto H^{2/3}$ [134]. Near room temperature and above, where the Dulong-Petit law applies, Gottschall *et al.* [65] indicated in 2019 that $\Delta T_{ad}^{max} \propto T_C^{1/3}$, although they did not state this explicitly. The $T_C^{1/3}$ relationship can be also deduced from the work by Oesterreicher *et al.*, in which they also assume that the lattice heat capacity is a constant [128]. **Figure 9** (a) shows $\Delta T_{ad}$ with respect to $T_C^{1/3}$ using the data from **Figure 7** (b) and (c), and the experimental data of $R$Al$_2$. As confirmed by the inset in **Figure 9** (a), the $T_C^{1/3}$ relationship holds near room temperature. Such a decrease of $\Delta T_{ad}^{max}$ with $T_C$ has been observed in the pure rare-earth metals Gd, Dy, and Tb, and the 5:4 compounds Tb$_5$Si$_4$ and Gd$_5$Si$_4$, as shown in **Figure 6**.

However, towards lower temperatures, where the Dulong-Petit law is no longer valid, and the lattice heat capacity can no longer be regarded as constant, the mean-field approach predicts that $\Delta T_{ad}^{max}$ increases with decreasing $T_C$, which is in accordance with the experimental data of $R$Al$_2$. Such an increase in $\Delta T_{ad}^{max}$ is a result of the sharply reduced lattice heat capacity. **Figure 9** (b) shows the molar lattice heat capacity of LuAl$_2$ and LaAl$_2$ using the Debye model, and the estimated lattice heat capacities at the Curie temperatures for the $R$Al$_2$ intermetallic series. As suggested by von Ranke et al. and Ribeiro et al. [34,59], we estimate the Debye temperatures of the $R$Al$_2$ intermetallic alloys with the Debye temperature of LaAl$_2$ and LuAl$_2$ using the following relationship:

$$T_D^{RAl_2}(T,n) = \frac{[14-n]T_D^{LaAl_2}(T) + nT_D^{LuAl_2}(T)}{14}, \quad (6)$$

where n is defined as the relative position of the rare-earth element in the rare-earth series. The volumetric lattice heat capacities of the $R$Al$_2$ at their Curie temperatures are then estimated using the estimated Debye temperatures. As seen from **Figure 9** (b), the lattice heat capacities of the $R$Al$_2$ at the $T_C$ decreases sharply: starting from GdAl$_2$, with a value close to 60 J K$^{-1}$ mol$^{-1}$, to ErAl$_2$, with a value of 0.24 J K$^{-1}$ mol$^{-1}$. Such a significant sharp drop in the lattice heat capacity happens for the $R$Al$_2$ series at about the $T_C$ of GdAl$_2$, which corresponds to the point where $\Delta T_{ad}^{max}$ starts to increase with decreasing $T_C$ in **Figure 9** (a).

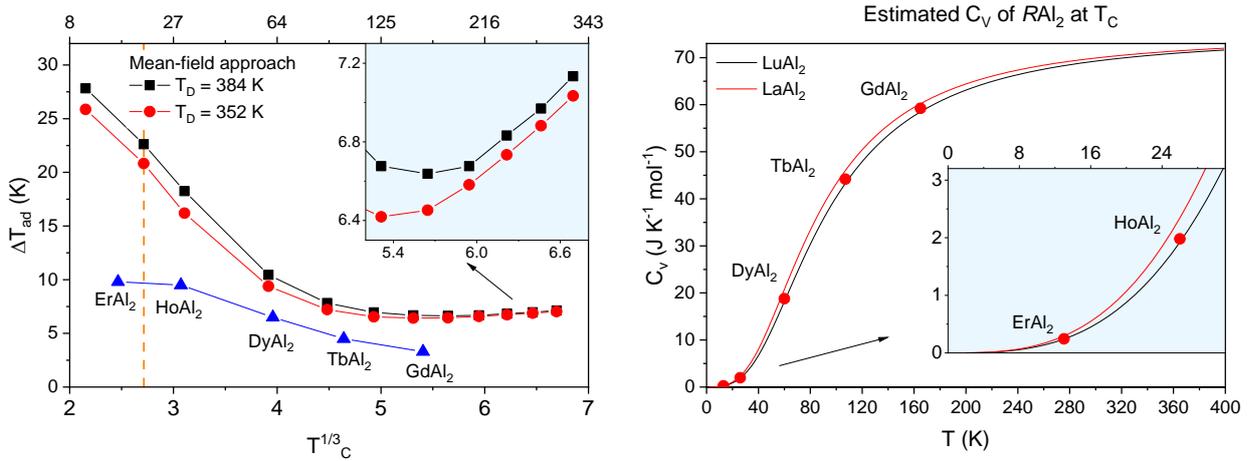

**Figure 9.** (a) Adiabatic temperature change as a function of $T_C^{1/3}$ using theoretical data (squares and circles) from **Figure 7** (b) and (c) as well as experimental data of $R$Al$_2$ (triangles) from **Figure 4** (e). The inset shows the enlarged area near room temperature. The vertical orange dashed line indicates the boiling point of H$_2$. (b) The molar lattice heat capacity of LuAl$_2$ and LaAl$_2$ and the estimated lattice heat capacity at $T_C$ of $R$Al$_2$ marked with red circles.

The mean-field approach can well explain the two trends, which are based on empirical findings from our literature review. The way that $T_C$ correlates with $\Delta S_m^{max}$ can be simplified to the negative exponential function $\Delta S_m^{max} \propto T_C^{-2/3}$, which holds true at least down to 20 K. Moreover, the sharp change in the lattice heat capacity leads to an increase of $\Delta T_{ad}^{max}$ with respect to the decreasing $T_C$ in the low-temperature range. This mean-field approach could be further developed to explore the field dependence of these two trends. In addition, the crystalline electric fields and the interactions between two different rare-earth ions when substitution is applied need to be considered if we are to perform the more accurate modeling [34], though it would complicate the whole process. Nevertheless, our mean-field approach does reveal a correlation between the Curie temperature and the maximum magnetocaloric effects vividly.

## 5. Summary
Our study of the Laves phases $R$Al$_2$ and $R$Ni$_2$ (excluding ErNi$_2$) revealed that both their $\Delta S_m^{max}$ and $\Delta T_{ad}^{max}$ increase with decreasing $T_C$. We then reviewed the literature for the quantities $\Delta S_m^{max}$, $\Delta T_{ad}^{max}$, and $T_C$ of rare-earth-based magnetocaloric materials, on the basis of which we summarized two trends for a intermetallic series: (1) $\Delta S_m^{max}$ increases with decreasing $T_C$; (2) $\Delta T_{ad}^{max}$ decreases with decreasing $T_C$ from room temperature, but then increases at cryogenic temperatures. Importantly, we discovered a noticeable, but unaddressed feature for the rare-earth-based magnetocaloric materials that magnetocaloric effect can be much stronger near 20 K, the boiling point of hydrogen.

We developed a mean-field approach to interpretate the correlations among $\Delta S_m^{max}$, $\Delta T_{ad}^{max}$, and $T_C$. The results provide support: we observed an increasing $\Delta S_m^{max}$ with decreasing $T_C$ and the non-monotonic dependence of $\Delta T_{ad}^{max}$ on $T_C$. The dependence of $\Delta S_m^{max}$ can be approximated to a simple form of $\Delta S_m^{max} \propto T_C^{-2/3}$, a negative exponential function that undergoes an ever-sharper increase towards an ever lower $T_C$. This gives a mathematical explanation as to why $\Delta S_m$ can exhibit large values around 20 K. In addition, by taking advantage of the similarities of the rare-earth elements, we examined this relationship using the Laves phases $R$Ni$_2$ and $R$Al$_2$. By combining the Debye model with the mean-field model, we show that the difference between the lattice heat capacity at cryogenic temperatures and that near room temperature is the key to understanding the non-monotonic dependence of $\Delta T_{ad}^{max}$ on $T_C$.

The dependence of $\Delta S_m$ and $\Delta T_{ad}$ on $T_C$ investigated here will help researchers to rapidly predict magnetocaloric properties, particularly for rare-earth-based magnetocaloric materials. It also helps researchers understand better how temperature acts on magnetocaloric effect, which not only can help design alloys for magnetocaloric liquefaction of hydrogen and other industrial gases, but also can benefit the search for new magnetocaloric materials.

**Declare of Interest**
The authors declare no conflict of interest.

**Acknowledgements**
The authors gratefully acknowledge financial support from the Helmholtz Association via the Helmholtz-RSF Joint Research Group (Project No. HRSF-0045), from the HLD at HZDR, member of the European Magnetic Field Laboratory (EMFL), from the DFG through the Würzburg-Dresden Cluster of Excellence on Complexity and Topology in Quantum Matter-ct.qmat (EXC 2147, Project ID 39085490), the CRC/TRR 270 (Project-ID 405553726) and the Project-ID 456263705, and from the ERC under the European Union's Horizon 2020 research and innovation program (Grant No. 743116, Cool Innov).

**Data Availability**
The data that support the findings of this study are available from the corresponding author upon reasonable request.